\documentclass[12pt]{iopart}

\usepackage[]{graphicx}

\begin{document}

\title[A study of magnetically-supported dc discharge in
cylindrical configuration]
{A study of magnetically-supported dc discharge in cylindrical
and inverted cylindrical configuration}

\author{O. Bilyk\footnote[3]{To
whom correspondence should be addressed (bilyk@mbox.troja.mff.cuni.cz)},
P. Kudrna, M. Hol\'\i k, A. Marek, M. Tich\'y}

\address{Charles University in Prague, Faculty of Mathematics and Physics,
V~Hole\v{s}ovi\v{c}k\'ach~2, 180~00 Prague~8, Czech Republic}

\begin{abstract}

We have investigated apparently
stochastic fluctuations of magnetically-supported dc discharge in
cylindrical coaxial
configuration. In the system the
electric field had radial direction while the magnetic field was applied axially.
The discharge vessel length was 12 centimetres. Working gas was typically
argon at pressure of several Pa, magnetic field 10-50 mT.

The contribution describes
experimental results - frequency and phase analysis
of the instabilities, which we detected in our experimental system in both
the conventional and inverted magnetron configurations.

We bring also  2-D PIC model of the dc discharge under
conditions, which can be achieved in the experimental apparatus.
The PIC model should answer questions, which originated from the
experimental study of the plasma parameters -- presented e.g. in
\cite{holik}.

\end{abstract}

\maketitle

\section{Introduction}

Magnetically-supported dc discharges in cylindrical symmetry (cylindrical magnetrons) are
used for deposition of high-temperature
superconducting materials, see e.g. \cite{lengl}, or materials with special
dielectric characteristics, see e.g. \cite{adam}. Understanding the behavior of
the dc discharge in such configuration
is therefore essential condition for the technological progress in this branch.
The magnetron systems with cylindrical symmetry
are relatively simple and thus the plasma processes in them can be simulated comparatively easily by
computer models. Measurements of plasma parameters in cylindrical magnetron
can thus be compared with theoretical
predictions and/or calculations used for their verification.

In this contribution results of the potential fluctuations measurements and 2-D PIC modelling are
brought. As we presented in former papers, e.g. \cite{holik}, by the study of axial
and radial dependence of the plasma parameters, the axial
inhomogeneities were observed, which couldn't be reproduced by
the 1-D PIC model. The 2-D model was then used to try to explain
the presence of the local minimum of the plasma density in the
middle of the discharge vessel. Further research has been done
in this direction and we bring new results of the calculations.
In connection with the differences between computer models and the experimental
results, importance is often attached to the plasma fluctuations.
To answer the question how big role play the fluctuations in our
system, we measured potential fluctuations with two parallel
probes and evaluated frequency spectra as well as a~correlation
of the signals from the both probes to check presence of
coherent modes \cite{martinez} in the discharge.

\begin{figure}[b]
\begin{center}
\scalebox{0.8}{
  \includegraphics{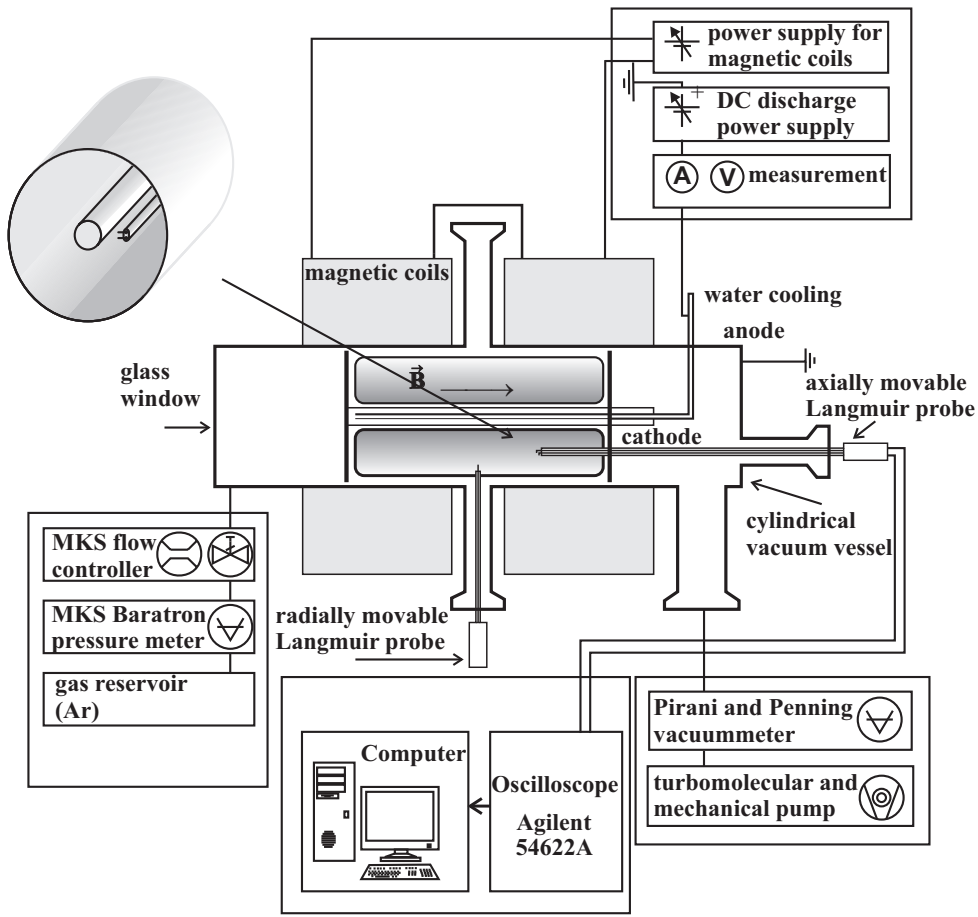}}
\end{center}
\caption{\label{magnetronold}Experimental set-up of the cylindrical magnetron
in coaxial configuration.}
\end{figure}

\section{Experimental}

\subsection{Experimental setup}

\begin{figure}
\begin{center}
  \scalebox{0.5}{
  \includegraphics*{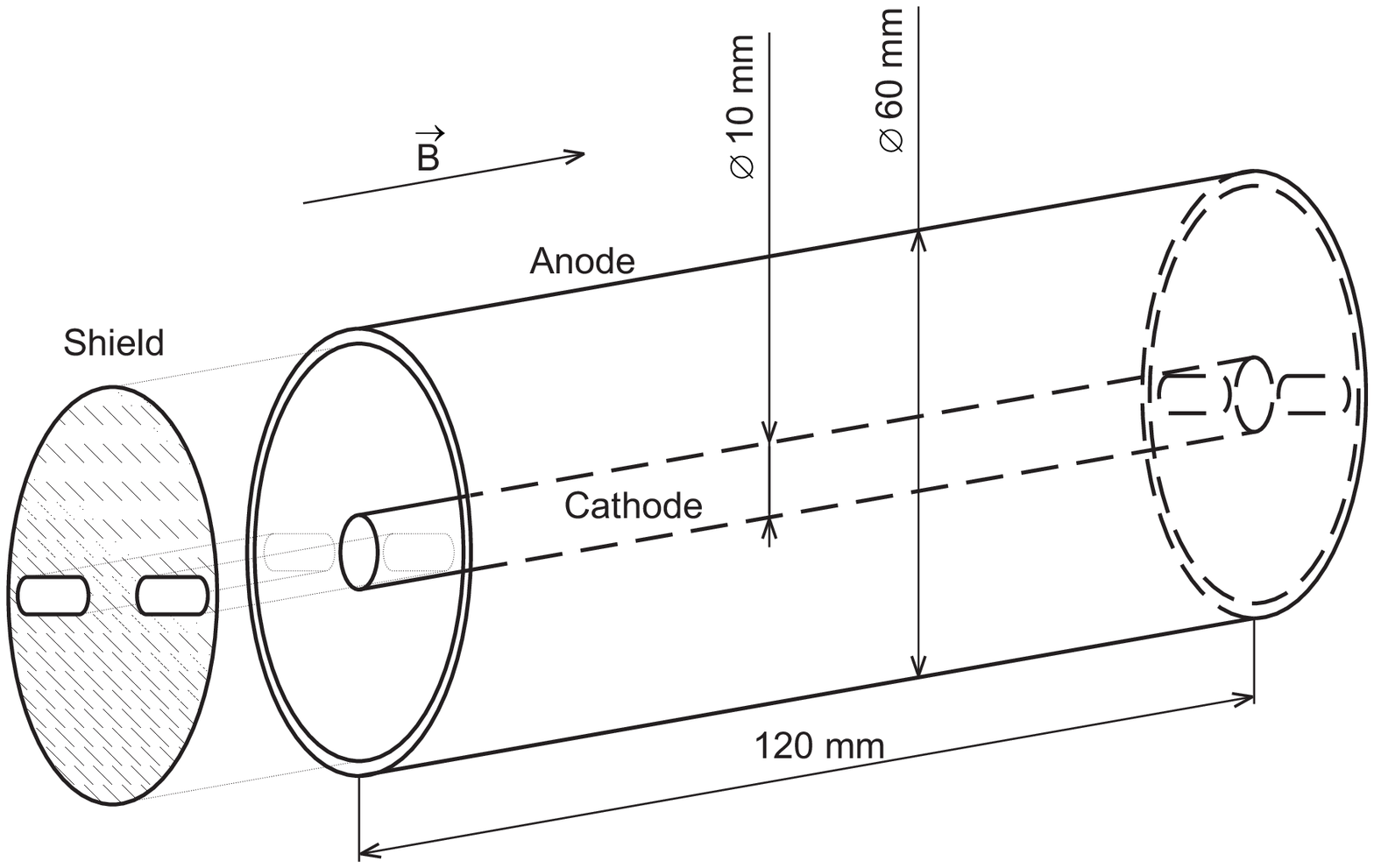}}
\end{center}
\caption{Geometric configuration of the magnetron electrodes -- the cathode is coaxially
placed in the middle of the vacuum vessel. In the inverted configuration the outer
electrode serves as a~grounded cathode and the anode is connected to a~positive dc power supply.}
\label{shape}
\end{figure}

The cylindrical magnetron in the so-called post discharge configuration
consists of cylindrical cathode mounted coaxially inside the anode.
The discharge volume is axially limited by means of two disc-shaped limiters,
which are connected to the cathode potential. Our device is schematically depicted in Fig.
\ref{magnetronold}, the electrodes configuration is shown in Fig. \ref{shape}.
The diameters
of the cathode and anode are $10~\mathrm{mm}$ and $60~\mathrm{mm}$, respectively. The length of the
discharge volume is $120~\mathrm{mm}$. The homogeneous magnetic field is created by
two coils and is in parallel with the common axis of the system. To prevent overheating
the coils and cathode, the system is cooled by water.

The apparatus is constructed as high vacuum. The pumping unit consists of the
combination of the turbomolecular and rotary pumps. The ultimate presure is
in the order of $10^{-3}~\mathrm{Pa}$. During the operation of the magnetron discharge the
working gas, in this case argon, slowly flows into the system and through the
valve reducing the pumping speed of the pumps leaves the system. The typical
flow rate is below $1~\mathrm{sccm}$ and is adjusted by means of the MKS flow controller
in order to keep the pressure in the discharge volume constant.

The system is equipped with several diagnostic ports, see Fig. \ref{magnetronold}. One of them can be
used to install a~radially movable Langmuir probe. Axially
movable probe can be installed using the port at the side of
the vacuum vessel. For measurements of the plasma potential
fluctuations the axial movable probe holder was used to enable
measuring with a~mean of two Langmuir probes placed as clear
from Fig. \ref{magnetronold}.

\subsection{Estimation of the power spectra}

The Langmuir probes were used in floating regime, i.e. without the applied bias voltage.
The floating potential
signal was sampled using the digital oscilloscope (Tektronix TDS 610) and samples
$h_n$ were transferred to the computer over the GPIB interface

\begin{equation}
h_n = h (n.\Delta t),
\label{hn}
\end{equation}
where $h(t)$ is the probe voltage with respect to the anode, $\Delta t$ is the
sampling interval and integer $n$ ranges from 0 to the number of samples $N-1$.
Then the discrete Fourier transform of the sampled data was calculated by
means of the FFT:

\begin{equation}
H_k = \sum_{n=0}^{N-1}h_n \mathrm{exp}\frac{-2\pi \mathrm{i} k n}{N}.
\label{Hk}
\end{equation}
The guess of the Fourier transform of the voltage $h(t)$ at the discrete
frequencies $k.\Delta f$ is given by

\begin{equation}
H(k.\Delta f) = \Delta t.H_k,
\label{guessofH}
\end{equation}
where $\Delta f = (N.\Delta t)^{-1}$. The power density is then

\begin{equation}
S(f) = \frac{1}{N.\Delta t}|{H(f)}|^2.
\label{power}
\end{equation}
In order to decrease the scatter of the calculated spectrum estimate several
realizations (typically 100) were averaged into the resulting power density curve.

\begin{figure}[h]
  \centerline{\scalebox{1.7}{\includegraphics*[5 mm,5 mm][97 mm,37 mm]{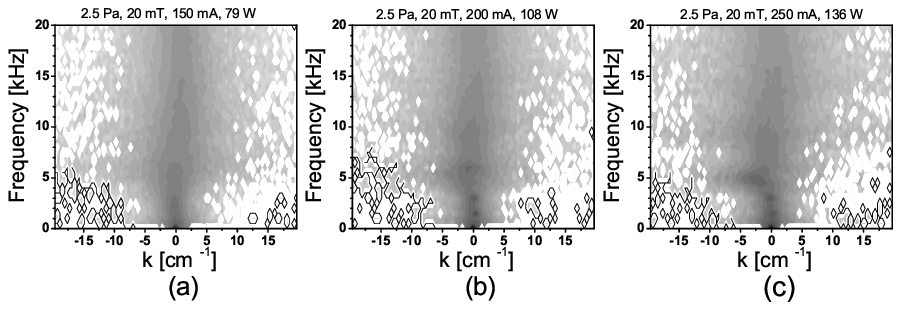}}}
  \centerline{\scalebox{1.7}{\includegraphics*[5 mm,5 mm][97 mm,37 mm]{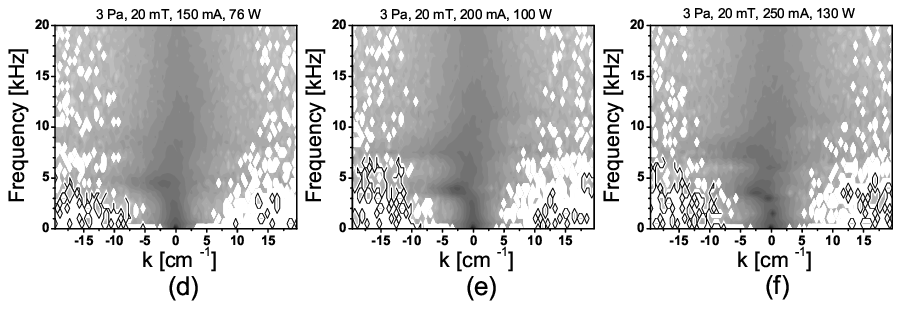}}}
 \caption{Frequency wave number histograms of the argon dc discharge in cylindrical magnetron
in conventional arrangement at pressure 2.5 Pa and 3 Pa and at magnetic field 20 mT.}
 \label{spectra}
\end{figure}

\subsection{Frequency vs. wave number spectra}

In order to analyze the wave behavior of the potential fluctuations
of the magnetron discharge the spectra derived from the fluctuations
measured simultaneously from two Langmuir probes were evaluated according
to the method \cite{martinez}. At first the wave number can be expressed
for each frequency from the phase shift between the signals at the two probes
and known distance $\Delta x$ between the probes:
\begin{equation}
k(f) = \frac{\mathrm{arg}H^{(2)}(f) - \mathrm{arg}H^{(1)}(f)}{\Delta x}.
\label{wavenum}
\end{equation}
Using the wave number and the average power density at two probes
$\frac{1}{2}[S^{(1)}(f)+ S^{(2)}(f)]$ the histogram $S(f,k)$ can be built.

In Figure \ref{spectra} there is shown an example of the
frequency wave number histograms, evaluated from the
simultaneous measurements with two probes at different discharge powers
in the non-inverted configuration.
The $z$-axis (power density axis) is showed as a~tone of grey
and darker areas at certain frequencies mean that the two
signals are correlated with given phase shift. There can be seen
certain coherent modes especially in Figures \ref{spectra}c through f.

The potential fluctuations of the discharge in the
inverted configuration were also investigated. However, although
the measurements were done within wide range of powers and
pressures, no observable coherent modes were detected in argon
discharge.

\section{PIC model}

Plasma modeling techniques can be divided into two basic families with
respect to the plasma description: codes using kinetic description and
codes using fluid description of plasma. The technique of our interest
-- Particle-In-Cell (PIC) technique e.g. \cite{birdsall, verbon, hammel}
-- belongs among the
codes that use kinetic description of plasma. In PIC technique there are
solved equations of motion for all charged particles in plasma. Computation
of mutual electrical forces is in the PIC approach transformed to the
solving of Poisson equation in the simulated region. A~great advantage
of this technique is that it is self-consistent. On the other hand
self-consistent approach is compensated by great computational effort
(especially in 2D and 3D simulations).

\begin{figure}
\begin{center}
\scalebox{0.5}{
  \includegraphics*{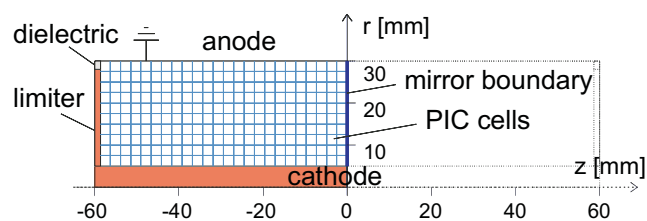}}
\end{center}
\caption{\label{cells}The electrodes configuration and the cells distribution
in the discharge area.}
\end{figure}

The electrostatic Particle-In-Cell computational scheme consists of the following steps:

1. Division of simulated region to cells.

2. Assignment of the charge of particles to the mesh points.

3. Computing of electric field.

4. Particles move in calculated field.

5. Jump to point 2.

\begin{figure}[b]
\begin{center}
  \scalebox{1.2}{
  \includegraphics*[0,35mm][150mm,80mm]{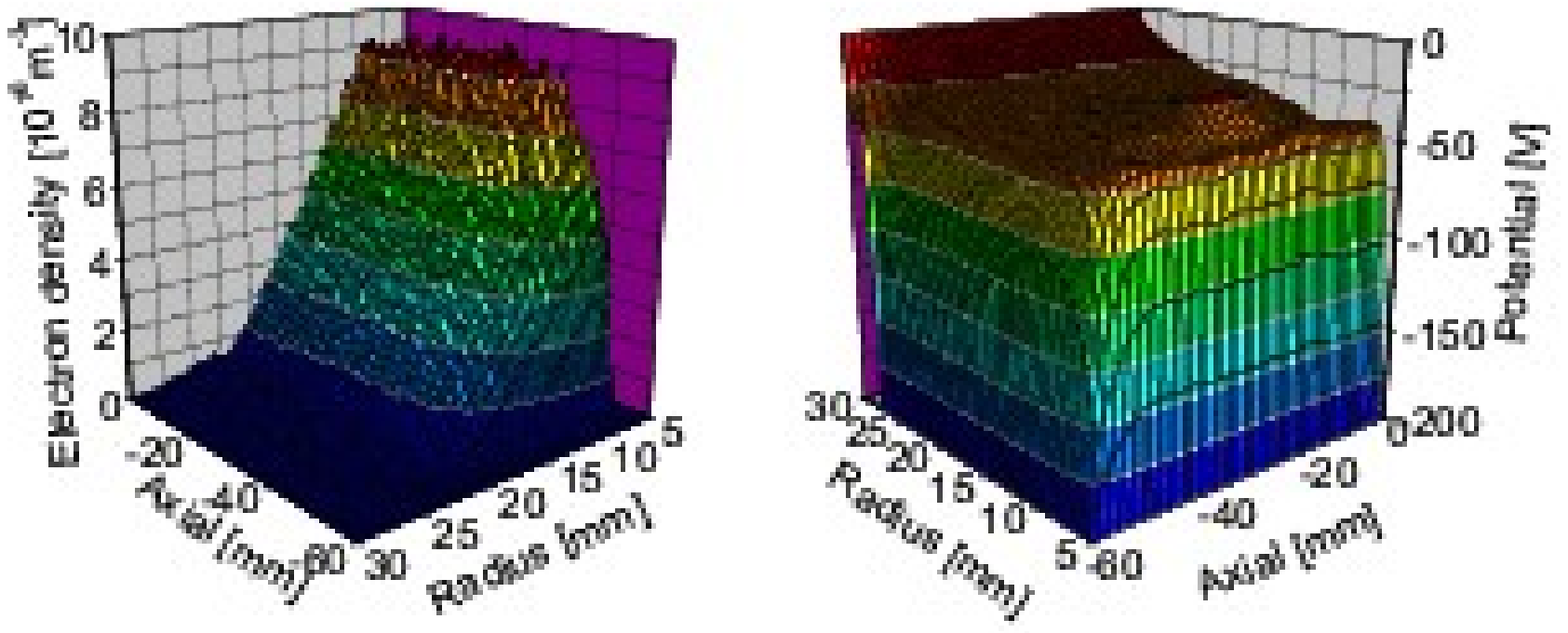}}
\end{center}
\caption{Results of the PIC modelling of the argon dc discharge in the
cylindrical magnetron apparatus.}
\label{pic}
\end{figure}

Although results of XOOPIC simulation presented here are in relatively
good agreement with the experiment there is a~problem with numerical
instability of the simulation -- with a~growing number of computational
particles during simulation. Hence, the computational model needs further
improvement. A~discharge current source may be implemented into the code
to help to solve this problem. In future work, we hope to bring with our
simulation model a~more detailed study of the discharge in cylindrical
magnetron and more precise comparison with experiment.

In Figure \ref{pic} there are shown the
results of the PIC modeling of the dc discharge in argon in the
cylindrical magnetron apparatus described above. The model was
done under the the following conditions: magnetic field strength
$20~\mathrm{mT}$, pressure $5~\mathrm{Pa}$, voltage $200~\mathrm{V}$,
time step $5~\mathrm{ps}$, simulation time $23.8~\mathrm{\mu
s}$. Grid size was 40 in axial direction and 100 in radial
direction.

\section*{Conclusion}

We measured potential fluctuations and evaluated the correlation
between two probes placed parallel in the discharge. The
coherent modes similar to those observed in \cite{martinez} were found under certain
conditions in the non-inverted cylindrical magnetron configuration.

The 2-D PIC model was used to calculate plasma parameters under
the same conditions, which can be established in the
experimental apparatus. The results of the electron density and
plasma potential were presented. The model results qualitatively
agree with experimental observations
\cite{holik,passoth,porokh1,porokh2}, further effort in modeling
is still needed.

\section*{Acknowledgments}

The work was financially supported by the Czech Science Foundation,
Grant No. 202/03/0827, 202/04/0360, 202/03/H162, by project COST action
527.70, by the Faculty of Mathematics and Physics of Charles University in Prague,
Research plan MSM 1132000002 and by EURATOM.


\section*{References}

\begin{thebibliography}{99}

\bibitem{holik} M. Hol\'{\i}k, O. Bilyk, A. Marek, P. Kudrna, J. F. Behnke, I. A. Porokhova,
Yu. B. Golubovskii, M. Tich\'y, 2004, {\itshape Contr. Plasma Phys.} {\bfseries 44}, No. 7-8, p. 613-618.

\bibitem{lengl} G. Lengl, P. Ziemann, F. Banhart, P. Walther, 2003, {\itshape Physica C}
\textbf{390}, p. 175.

\bibitem{adam} M. Adam, D. Fuchs, R. Schneider, 2002, {\itshape Physica C} \textbf{372}, p. 504.

\bibitem{martinez} E. Martines, et.al., 2001, {\itshape Phys. of Plasmas} {\bfseries 8}
6, 3042.

\bibitem{birdsall} C.K. Birdsall, A.B. Langdon, Plasma physics via computer simulation, Mc Graw-Hill 1995, Adam-Hilger 1991.

\bibitem{verbon} J. Verboncoeur, A.B. Langdon and N.T. Gladd, An Object-Oriented Electromagnetic PIC
Code, 1995, {\itshape Comp. Phys. Comm.} {\bfseries 87}, p. 199-211.

\bibitem{hammel} J.P. Hammel, J. Verboncoeur, DC Discharge studies using
PIC-MCC, 2004, http://ptsg.eecs.berkeley.edu/\~jhammel/report.pdf.

\bibitem{passoth} E. Passoth, P. Kudrna, C. Csambal, J.F. Behnke, M. Tichy,
V. Helbig, 1997, {\itshape J. Phys. D: Appl. Phys.} {\bfseries 30} p. 1763-1777.

\bibitem{porokh1} I. A. Porokhova, Yu. B. Golubovskii, J. Bretagne, M. Tichy,
and J.F. Behnke, 2001, {\itshape Physical Review E} {\bfseries 63}.

\bibitem{porokh2} I. A. Porokhova, Yu. B. Golubovskii, M. Holik, P. Kudrna,
M. Tichy, C. Wilke, and J. F. Behnke, 2003, {\itshape Physical Review E} {\bfseries 68}.

\end{thebibliography}
\end{document}